%% file: timedomain_OST4K.tex
\documentclass[
  aip,
  amsmath,
  amssymb,
  reprint,
]{revtex4-1}

\usepackage{graphicx}
\usepackage[dvipsnames]{xcolor}
\usepackage[urlcolor=blue, hyperindex, colorlinks, bookmarks=true]{hyperref}
\usepackage{dcolumn}% Align table columns on decimal point
\usepackage{bm}% bold math

\newcommand*{\m}[0]{\ensuremath{\mathbf{m}}}

\newcommand*{\h}[0]{\ensuremath{\mathbf{h}}}
\newcommand*{\heff}[0]{\ensuremath{\mathbf{h}_{\mathrm{eff}}}}
\newcommand*{\han}[0]{\ensuremath{\mathbf{h}_{\mathrm{an}}}}
\newcommand*{\gam}[1]{\ensuremath{\mathbf{\Gamma}_{\mathrm{#1}}}}

\newcommand*{\nstt}[0]{\ensuremath{\mathbf{n}_\text{stt}}}
\newcommand*{\deriv}[2]{\ensuremath{\frac{\text{d} #1}{\text{d} #2}}}

\newcommand{\units}[1]{\,\mathrm{#1}}

\begin{document}

\title{Coherent spin-transfer precession switching in orthogonal spin-torque devices}
\author{G. E. Rowlands}
\email{graham.rowlands@raytheon.com}
\affiliation{Raytheon BBN Technologies, Cambridge, MA 02138, USA}
\author{C. A. Ryan}
\affiliation{Raytheon BBN Technologies, Cambridge, MA 02138, USA}
\author{L. Ye}
\affiliation{Center for Quantum Phenomena, Department of Physics, New York University, New York, NY 10003, USA}
\author{L. Rehm}
\affiliation{Center for Quantum Phenomena, Department of Physics, New York University, New York, NY 10003, USA}
\author{D. Pinna}
\affiliation{Center for Quantum Phenomena, Department of Physics, New York University, New York, NY 10003, USA}
\author{A. D. Kent}
\email{andy.kent@nyu.edu}
\affiliation{Center for Quantum Phenomena, Department of Physics, New York University, New York, NY 10003, USA}
\author{T. A. Ohki}
\email{thomas.ohki@raytheon.com}
\affiliation{Raytheon BBN Technologies, Cambridge, MA 02138, USA}

\date{\today}

\begin{abstract}

We present experimental results and macrospin simulations of the switching
characteristics of orthogonal spin-transfer devices incorporating an
out-of-plane magnetized polarizing layer and an in-plane magnetized spin valve
device at cryogenic temperatures. At $T \approx 4\units{K}$ we demonstrate: high
speed deterministic switching at short pulse lengths --- down to $100\units{ps}$ ---
with sufficient measurement statistics to establish a switching error rate of
$10^{-5}$; coherent precessional switching at longer times; and observe ensemble
decoherence effects at even longer times. Finite temperature macrospin models
model the precessional switching well but fail to fully reproduce all the
decoherence and switching error behaviour.

\end{abstract}

% \pacs{Valid PACS appear here}

\maketitle

Spin-transfer devices that operate at low temperature are of interest for
applications that require a cryogenic memory, such as Josephson junction based
logic circuits \cite{Holmes2013}. A requirement for this application is high
speed operation with relatively low energy dissipation. Conventional spin
transfer torque (STT) devices, such as those being developed as commercial room
temperature memories, may not be suitable for low temperature operation if the
switching is thermally activated and may not take advantage of the lower power
possible at low temperatures. A conventional STT device consists of two thin
magnetic layers separated by a non-magnetic layer, with the memory states being
the layer magnetizations aligned either parallel or antiparallel. However, the
initial spin-transfer torque in these states is vanishingly small, and a spin
polarized current amplifies thermal fluctuations of the magnetization, leading
to nanosecond incubation delays for switching and stochastic switching
characteristics \cite{Brataas2012, Devolder2008, Liu2014}. At low temperature
this mechanism of STT writing should be even slower and less energy efficient.

\begin{figure}[t]
  \includegraphics[width=\columnwidth]{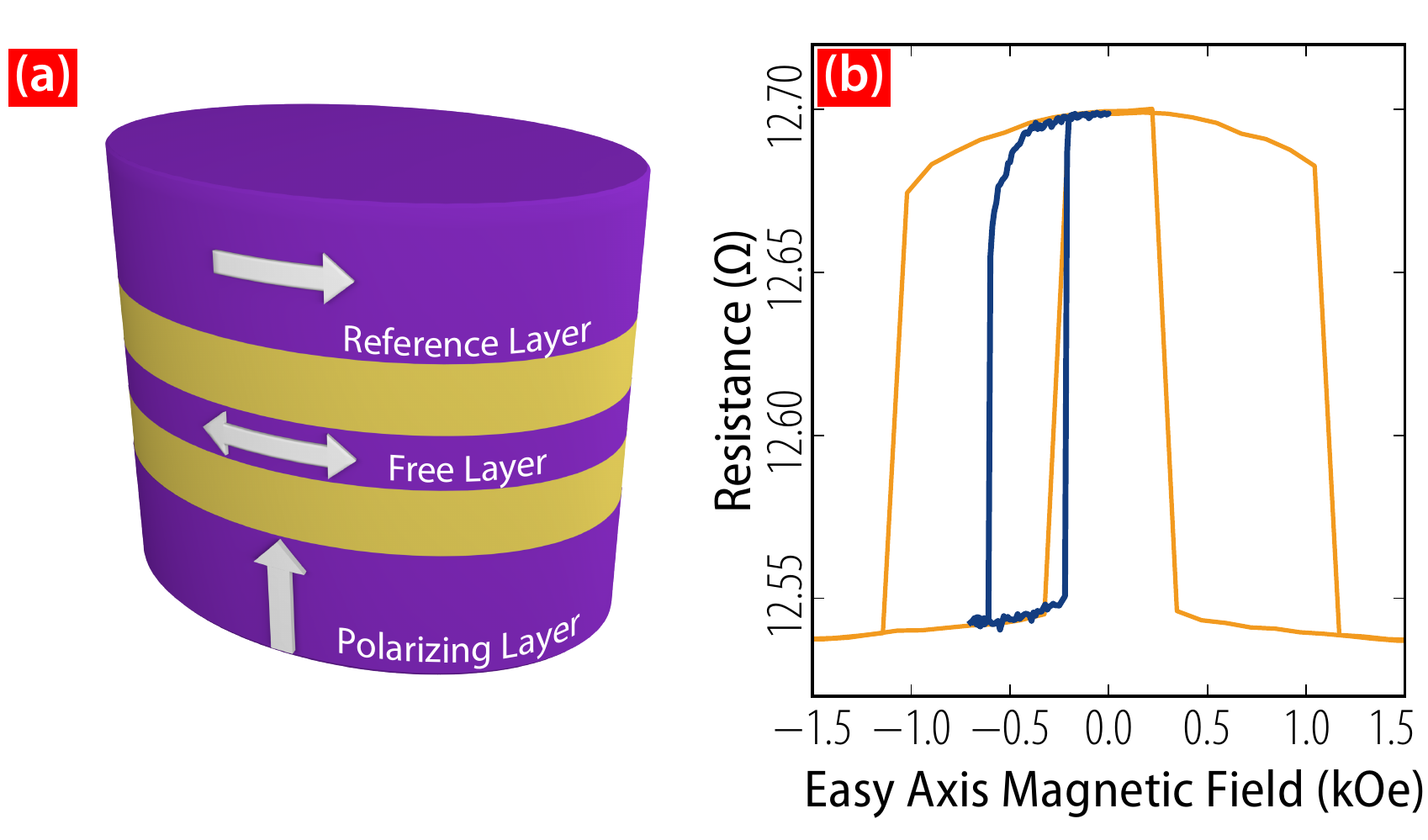}
  \caption{\label{fig:schematic}(a) Schematic of an OST device, with the
  equilibrium magnetization directions of the free, fixed, and polarizing layers
  indicated. (b) Major (orange) and minor (blue) hysteresis loops of a device at $T = 4\units{K}$.}
\end{figure}

An orthogonal spin-transfer (OST) device overcomes this limitation by having a
spin-polarizing layer aligned perpendicular to the free layer that provides a
large spin-torque the moment a current is applied \cite{Kent2004}. This
perpendicular polarizer induces precessional magnetization dynamics, as it
forces the free layer magnetization out of the film plane leading to coherent
precessional motion of the magnetization about the film normal, the magnet's
hard magnetic axis \cite{Houssameddine2007}. Experiments at room temperature
have demonstrated both fast ($<1\units{ns}$) switching as well as the excitation of
precessional magnetization dynamics \cite{Lee2009, Papusoi2009, Beaujour2009,
Liu2010, Rowlands2011, Liu2012} while recent experiments at low temperature have demonstrated
even faster ($<100\units{ps}$) and deterministic switching in OST-devices
\cite{Park2013,Ye2014}. These experiments were conducted over a relatively limited
range of applied currents and pulse durations, and thus could not probe the
coherent magnetization dynamics expected at low temperature.

Here we study switching in OST spin-value based devices at $T\approx 3\units{K}$
with an improved experimental setup focused on minimizing thermal noise at the
device increasing the data taking rate. This enables us to measure a range of
pulse amplitudes and durations to observe phase diagrams of coherent
magnetization dynamics. We also demonstrate deterministic high speed switching
for pulse lengths down to $100\units{ps}$ and establish bounds on the switching
error rates not limited by measurement statistics. We model our results with a
finite-temperature stochastic Landau-Lifshitz-Gilbert-Slonczewski (LLGS) model
\cite{Slonczewski1996} and find that even at cryogenic temperatures, thermal
noise or micro-magnetic effects can lead to an incoherent decay in the switching
probabilities.

\begin{figure*}[tb!]
    \centering
    \includegraphics[width=0.8\textwidth]{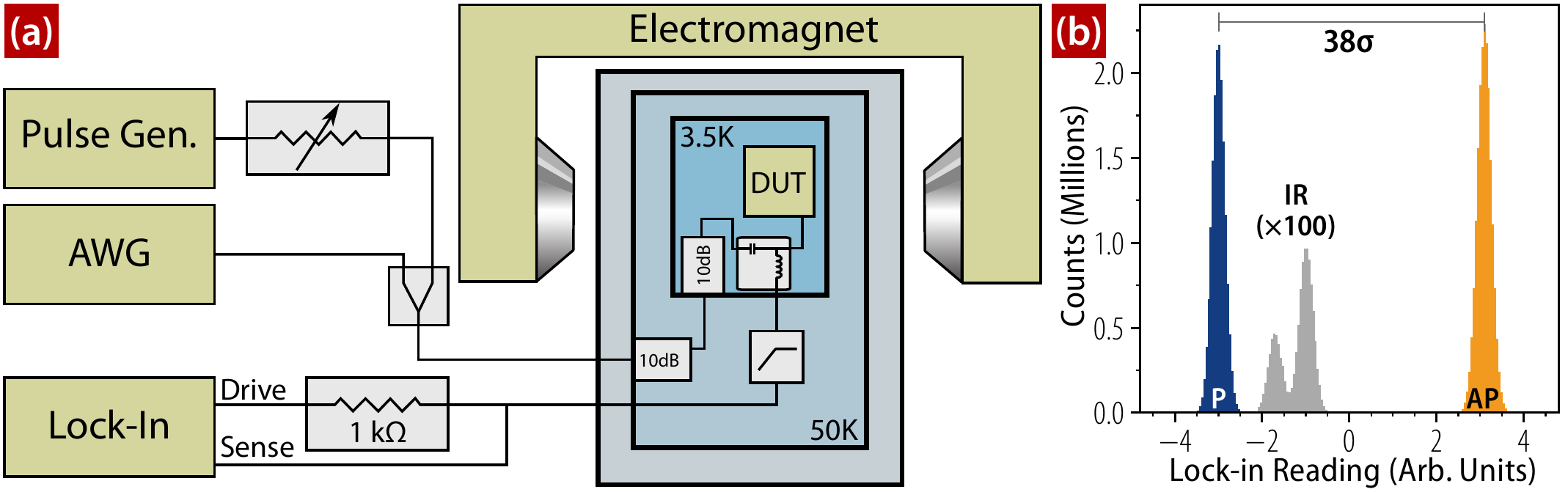}
    \caption{\label{fig:experimental-schematic} (a) Block diagram of the
    measurement apparatus.  High-speed pulses from either the pulse generator or
    AWG are capacitively coupled into the single-port device under test (DUT)
    via a bias-T. The low-speed arm of the bias-T is used by the lock-in
    amplifier to measure the device state. (b) Histograms of the sample's
    voltage states from a typical run of 35.6 million shots. The separation is
    shown in terms of the average standard deviation $\sigma$ of the P and AP
    states. The intermediate resistance (IR) state is shown magnified by a
    factor of 100.}
    \end{figure*}

The OST devices under study consist of a CoFeB(3) magnetic free layer (FL), a
CoFeB(12) reference layer (RL), and a [Co(0.3)/Pd(0.7)]$_2$/[Co(0.15)/Ni(0.6)]$_3$
perpendicularly magnetized spin-polarizing layer (PL) arranged into a full stack of
$||$PL/Cu(10)/FL/Cu(10)/RL$||$ as shown in Figure~\ref{fig:schematic}(a). All
dimensions are given in nm.  Nanopillars of various shapes and aspect ratios
were fabricated using e-beam lithography and ion-milling. Here we present
results for devices with a $50\units{nm} \times 100\units{nm}$ elliptical
cross-section, whose FLs have shape anisotropy that defines a magnetic easy
axis in the film plane along the long axis of ellipse. Shape anisotropy also
sets the magnetization direction of the thicker RL. The major and minor
hysteresis loops of one such device are shown in Figure~\ref{fig:schematic}(b).
A clear offset of the minor loop is observed because of the uncompensated dipole
field from the RL.

Sample chips are mounted and wirebonded in a custom package designed to support
microwave signals. The package is mounted on the cold-head of Gifford-McMahon
cryocooler (Sumitomo RDK-101D) with a base temperature of $\approx 4\units{K}$
when loaded with coaxial lines. Switching pulses are provided by either a pulse
generator (Picosecond Pulse Labs 10,070A) or an AWG (Keysight M8190A) that are
combined and then capacitively coupled to the device via a bias-T (Picosecond
Pulse Labs 5575A) mounted at the cold head. The high speed line has cryogenic
attenuators at both the $50\units{K}$ and $3\units{K}$ stages to thermalize the
center conductor and attenuate thermal noise from higher temperature stages. To
provide additional pulse amplitude resolution for the pulse generator (above the
$1\units{dB}$ resolution of the internal step attenuator) we use a voltage controlled
variable attenuator (RFMD RFSA2113 evaluation board). The DC coupled port of the
bias-T is used to apply current bias and make resistance measurements using a
lock-in amplifier (SRS 865) operated at a $1\units{MHz}$ baseband. The
measurement line is low-pass filtered with a custom ECCOSORB low-pass
filter~\cite{Santavicca2008} at the $3\units{K}$ stage again to suppress thermal
noise from room temperature. The external magnetic field is applied by a
room-temperature electromagnet. A block diagram of the setup is shown in
Figure~\ref{fig:experimental-schematic}(a).

Switching studies are performed with a bias field set in the center of the minor
hysteresis loop of Figure~\ref{fig:schematic}(b). Since precessional switching
can be induced irrespective of polarity, we use the AWG to supply a
fixed-polarity ``reset'' pulse whose width and amplitude produce high
probability switching regardless of whether the system starts in the
antiparallel (AP) or parallel (P) configuration of the FL and RL. Reset pulses
are applied every second shot followed by the switching pulse applied using the
pulse generator. This procedure allows us measure both AP$\to$P and P$\to$AP
transitions in the same data run. To maximize the data acquisition rate, the
experiment is sequenced by the AWG and is continuously streamed at rate of
$\approx 10\units{kHz}$ until the desired switching statistics are achieved. The
voltages from the lock-in were clustered into two or three clusters using a
kmeans algorithm~\cite{Macqueen1967} (see
Figure~\ref{fig:experimental-schematic}(b)). With a sufficient settling time for
the lock-in the distributions are approximately Gaussian --- insomuch as they are
visually indistinguishable from a normal distribution despite failing the
Anderson-Darling test for normality --- and well separated. Taking the Gaussian
assumption the probability of misclassification of the state is infinitesimally
small at $< 10^{-82}$ and thus the reported switching error rates are intrinsic
to the device. The third cluster was used for when we occasionally saw an
unresponsive intermediate resistance (IR) state~\cite{Ye2015}, most commonly after the
application of high amplitude switching pulses. Typically, the device would
remain stuck in this IR state for a short period of time but occasionally would
need to be forced out with a magnetic field sweep. We hypothesize that this
state arises from a non-uniform magnetization state in the FL and/or RL that
decreases the switching efficiency of STT induced reversal.

\begin{figure}
\includegraphics[width=\columnwidth]{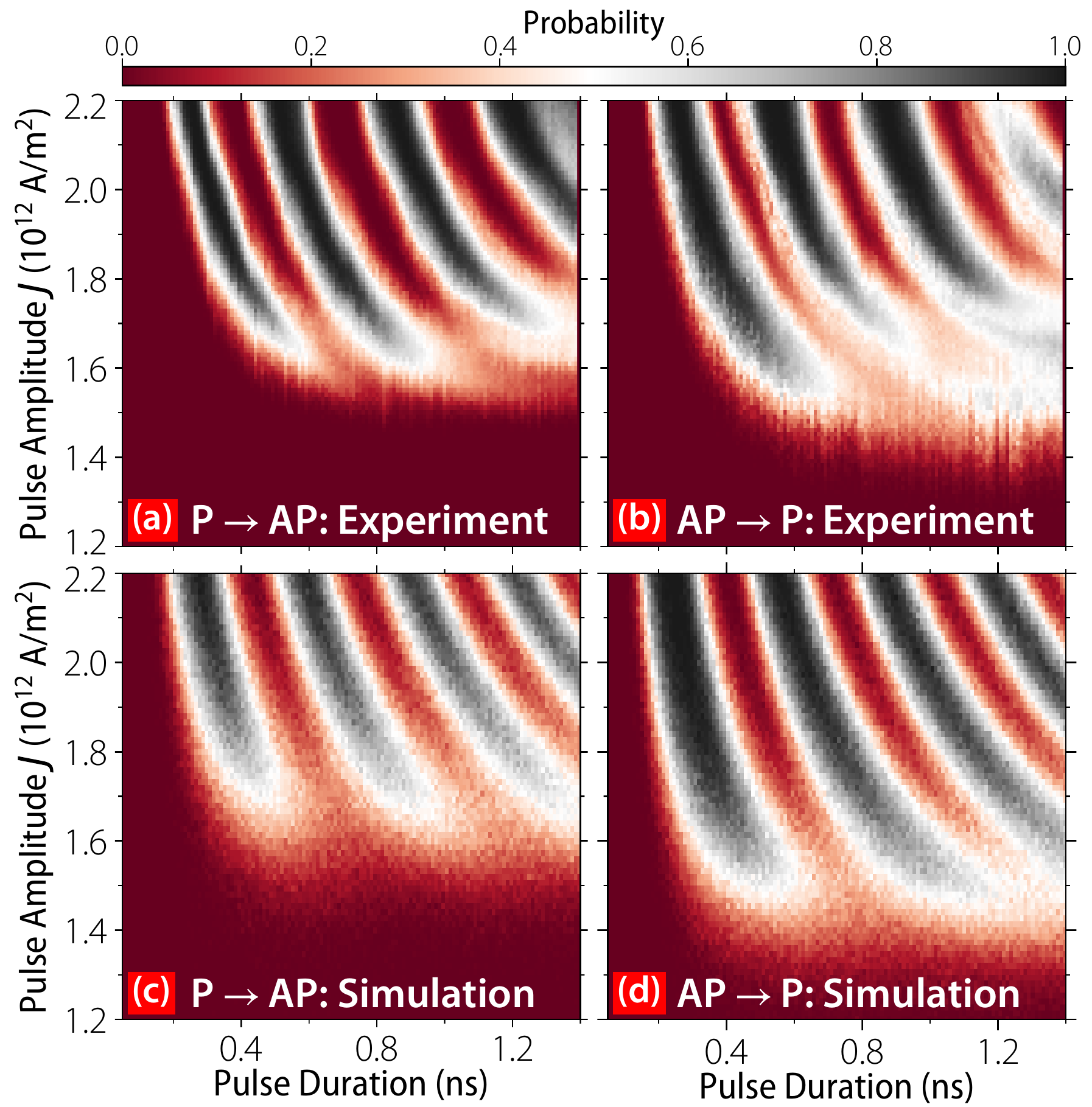}
\caption{\label{fig:PhaseDiagrams} Comparison of $T=4$ K experimental (a,b) and
$T=60$ K simulated (c,d) switching phase diagrams for (a,c) P$\to$AP and (b,d)
AP$\to$P switching polarities. Each pixel represents an estimate of the switching
probability from on average 2048 shots.}
\end{figure}

We apply this measurement procedure over a range of pulse amplitudes and
durations in order to build up a phase diagram for precessional switching. These
results are shown in Figs.~\ref{fig:PhaseDiagrams}(a-b), where each pixel
represents the switching probability calculated as the mean of the beta
distribution for the $\approx 2048$ switching attempts in each direction (we
take 4096 shots for each pair of pulse amplitude and duration settings and with the
reset pulse every second shot the initial state distribution is
$\mathrm{P}:\mathrm{AP} = 50.9:49.1 \pm 0.05$). In both AP$\to$P and P$\to$AP
polarities, the sample undergoes three full probability oscillations with a
period of approximately $400\units{ps}$. For longer pulses the switching
probability does not recover to 100\%, and for longer pulses yet (not shown) the
sample can occasionally become stuck in the intermediate resistance state
mentioned above.

To better understand these results we perform finite temperature simulations of
macrospin model of Eq.~\ref{LLGS} using our parallel, GPU enabled,
\texttt{macrospin\_gpu} package~\cite{MacrospinGPU:2017}. The simulated dynamics
are described by the LLGS equation:
\begin{equation} \label{LLGS}
  \deriv{\m}{\tau}=-\gam{llg}+\gam{th}+\gam{stt}.
\end{equation}
where $\m=\mathbf{M}/M_s$ is the FL magnetization unit vector, $\gam{llg}$ the
deterministic LLG torque, $\gam{th}$ the thermal torque, and $\gam{stt}$ the
STT. The LLG torque, $\gam{llg}=-\m\times\heff-\alpha\m\times(\m\times\heff)$,
is given in terms of the effective field $\heff=\frac{-1}{\mu_0
M_s^2V}\nabla_{\m}U(\m)$ for FL volume $V$ and damping constant $\alpha$. Time
$t$ in Eq.~\ref{LLGS} has been normalized by the precession frequency so that
$\tau=\gamma \mu_0 M_st$, where $\gamma$ is the gyromagnetic ratio. The thermal
torque $\gam{th}$ is induced by a Gaussian distributed random field
$\mathbf{h}_{\mathrm{th}}$~\cite{Brown1963}. Although this is a standard
assumption we note its applicability to the present regime of cryogenic
temperatures and sub-nanosecond switching times is not established. The combined
spin-torque contributions from both the out-of-plane PL and in-plane RL can be
described in terms of effective spin-polarization vector $\nstt$:
\begin{align} \label{STT}
  \gam{stt} &= \tilde{I}\m\times(\m\times\nstt) \nonumber\\
  \nstt &= \mathcal{P}_R\eta(\Lambda_R, \m_x)\mathbf{\hat{x}} + \mathcal{P}_P\eta(\Lambda_P, \m_z)\mathbf{\hat{z}} \\
  \eta(\Lambda, \cos\theta) &= \frac{2\Lambda^2}{(\Lambda^2+1) + (\Lambda^2-1)\cos\theta}.
\end{align}
Here $\mathcal{P}_{R,P}$ and $\Lambda_{R,P}$ are the spin-torque polarization
and asymmetry parameters~\cite{Slonczewski2002}, respectively, $\eta(\Lambda,
\cos\theta)$ encodes the angular dependence of the spin torque for an angle
$\theta$ between the spin torque polarization and FL, and
$\tilde{I}=(\hbar/2e)I/(\mu_0M_s^2V)$ is the normalized applied current.

As in the experiments we simulate the entire amplitude-duration phase diagram,
shown in Figs.~\ref{fig:PhaseDiagrams}(c-d), where each pixel gives the
switching probability for an ensemble of 512 macrospins subject to different
realizations of the thermal field. We take $M_s = 1200 \times 10^3\units{A/m}$,
$\alpha = 0.06$, $\mathcal{P}_{R/P} = 0.03/0.05$, $\Lambda_{R/P} = 1.5/1.0$. The
shape anisotropy is treated as the combination of two uniaxial contributions:
an out-of-plane hard-axis demagnetizing field $\h_\text{d} = 4 \pi M_s$ and an
in-plane easy-axis field $\han = 100$ Oe. The dipolar field from
the reference layer is assumed to be cancelled by the external field and is
omitted in our simulations. The ensemble is allowed to thermalize in the AP or
P state before a current pulse (rise/fall time of $65/110\units{ps}$) is
applied. The simulations reproduce the shape and periodicity of the probability
oscillations. The disparity in critical current densities $J_c$ for different switching
polarities, as well as the slightly increased AP$\to$P switching speed is
caused by the influence of STT from the in-plane RL~\cite{Ye2015}. For negative pulses (not
shown), we find reduced switching probabilities that are indicative of a
non-zero $\lambda_R$ \cite{Park2013}.

\begin{figure}
  \includegraphics[width=\columnwidth]{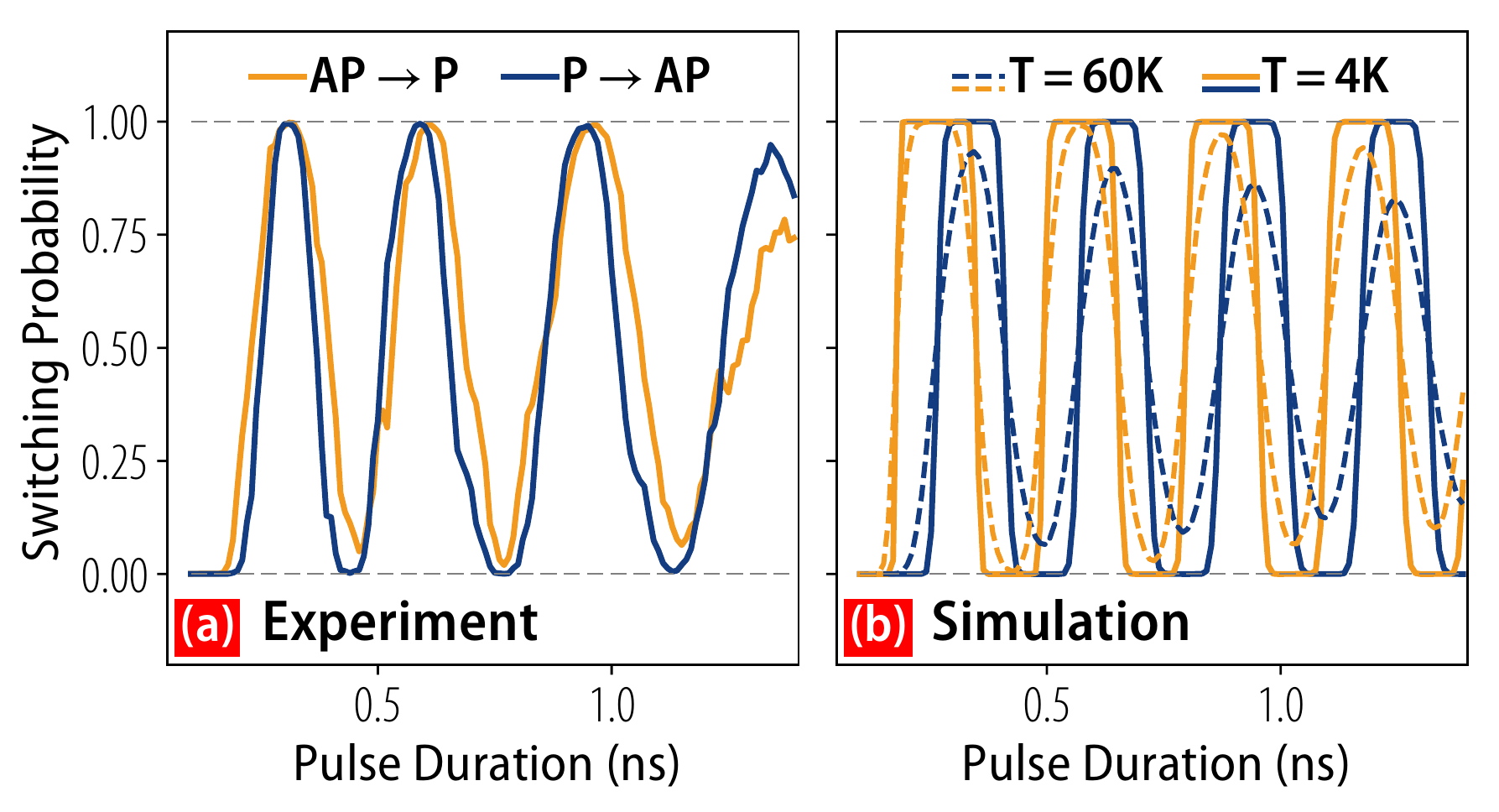}
  \caption{\label{fig:probability_cuts} (a) Cuts of the experimental phase
  diagrams taken at $J=2.0 \times 10^{12}\units{A/m^2}$.(b) Cuts of the
  simulated phase diagrams at $J=0.52 \times 10^{12}\units{A/m^2}$. Data are
  shown for $T=4\units{K}$ (solid lines) and $T=60\units{K}$ (dashed lines).
  AP$\to$P (orange) and P$\to$AP (blue) switching polarities are shown in both
  in both (a) and (b).}
\end{figure}

The simulations of Figs.~\ref{fig:PhaseDiagrams}(c-d) are performed at
$T=60\units{K}$ in order to produce a broadening of the switching bands
consistent with the experimental data. Comparing constant pulse-amplitude slices of the
simulated and experimental phase diagrams, as shown in Figs.
\ref{fig:probability_cuts}(a,b), reveals some distinct differences. At
$T\approx 4\units{K}$ the experimental probability oscillations exhibit a sinusoidal
behavior before a precipitous decline in probability upon the final oscillation.
Meanwhile, the simulated $T=4\units{K}$ oscillations exhibit wide high-probability
bands with a minimal amount of rounding. At $T=60\units{K}$ the simulations show
a gradual decoherence of the ensemble resulting from dephasing along the
out-of-plane switching trajectories.~\cite{Pinna2014a} This behavior is not observed in the
experimental data, and we hypothesize that a micromagnetic instability (perhaps
in the form of domain nucleation) is responsible for the abrupt departure from
full-probability oscillations. The more complicated sub-structure of the
experimental switching phase diagrams near threshold may also result from
micromagnetic considerations.

\begin{figure}
    \includegraphics[width=0.7\columnwidth]{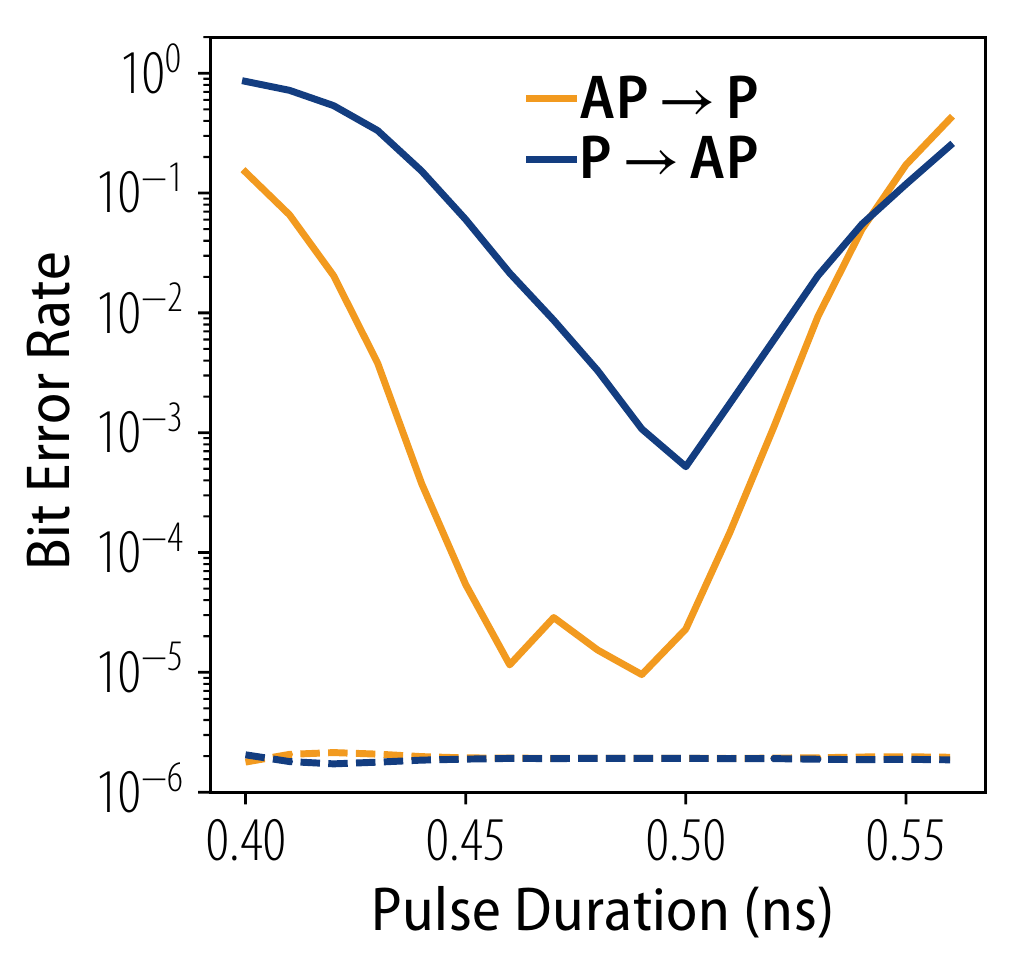}
    \caption{\label{fig:error_rate} The error rates for AP$\to$P (orange) and
    P$\to$AP (blue) switching for $J=2.2 \times 10^{12}\units{A/m^2}$ in the
    vicinity of the second probability oscillation. Statistical limits taken
    from the beta distribution (dotted lines) are shown independently for each
    polarity.}
\end{figure}

The experimental AP$\to$P and P$\to$AP transition probabilities in
Fig. \ref{fig:probability_cuts}(a) are synchronized such that they are maximized
for the same pulse durations, in contrast to the behavior of the macrospin
simulations in Fig. \ref{fig:probability_cuts}(b).
This has important implications for memory write circuitry, which would
suffer a substantial increase in complexity were it required to generate
different pulses of different amplitudes. We explore the bit error rates (BER)
near the maximum of the second probability oscillation, as shown in Figure
\ref{fig:error_rate}. We find that AP$\to$P switching reaches $10^{-5}$ error
rates over a fairly broad window of pulse widths. For P$\to$AP switching, there
is a comparatively narrow window where switching reaches a BER of $10^{-3}$.

In conclusion, we have measured high-resolution switching phase diagrams for
orthogonal spin-torque devices that reveal precessional reversal. Finite
temperature macrospin simulations reproduce many of the qualitative features of
this reversal, and allow us to identify the clear influence of STT from both the
PL and RL. Simulations cannot reproduce the sinusoidal probability oscillations
without introducing strong dephasing, nor can they account for the coincidence
between AP$\to$P and P$\to$AP switching probability maxima. Further work,
particularly in the area of micromagnetic simulations, is required to understand
the origin of these features.

We thank Robert Buhrman for many productive discussions. Measurements are
performed using the Auspex package~\cite{Auspex:2017}. The macrospin simulations
were performed using the macrospin\_gpu package~\cite{MacrospinGPU:2017}. Data
analysis was performed in Julia~\cite{Bezanson2014} with k-means clustering from
the Clustering.jl package\cite{Clustering.jl}. The figures were made using
matplotlib~\cite{Hunter2007}. The research is based on work supported by the Office of the Director of National Intelligence (ODNI), Intelligence Advanced Research Projects Activity (IARPA), via contract W911NF-14-C0089. The views and conclusions contained herein are those of the authors and should not be interpreted as necessarily representing the official policies or endorsements, either expressed or implied, of the ODNI, IARPA, or the U.S. Government. The U.S. Government is authorized to reproduce and distribute reprints for Governmental purposes notwithstanding any copyright annotation thereon. This document does not contain technology or technical data controlled under either the U.S. International Traffic in Arms Regulations or the U.S. Export Administration Regulations.

% \bibliography{BibCOST}
\input{timedomain_OST4K.bbl}
\end{document}